\RequirePackage{fix-cm}
\documentclass[smallextended]{svjour3}       % onecolumn (second format)
\smartqed  % flush right qed marks, e.g. at end of proof
\usepackage{graphicx}
\usepackage{subfigure}
\usepackage[numbers]{natbib}
\begin{document}

\title{Shape optimization of active and passive drag-reducing devices on a D-shaped bluff body}

\titlerunning{Shape optimizations on a D-shaped bluff body}        % if too long for running head

\author{Richard Semaan}


\institute{Institute of Fluid Mechanics, Technical University of Braunschweig \at
              Hermann-Blenk-Str. 37, 38108, Braunschweig, Germany \\
              Tel.: +49-531-391-94258\\
              Fax: +49-531-391-94254\\
              \email{r.semaan@tu-bs.de}           %  \\
}

\date{Received: date / Accepted: date}

\maketitle

\begin{abstract}
Shape optimization of an active and a passive drag-reducing device on a two-dimensional D-shaped bluff body is performed. 
The two devices are: Coanda actuator, and randomly-shaped trailing-edge flap. 
The optimization sequence is performed by coupling the genetic algorithm software DAKOTA to the mesh generator Pointwise and to the CFD solver OpenFOAM.
For the the active device the cost functional is the power ratio, whereas for the passive device it is the drag coefficient.
The optimization leads to total power savings of $\approx 70\%$ for the optimal Coanda actuator, and a 40\% drag reduction for the optimal flap. 
This reduction is mainly achieved through streamlining the base flow and suppressing the vortex shedding. 
The addition of either an active or a passive device creates two additional smaller recirculation regions in the base cavity that shifts the larger 
recirculation region away from the body and increases the base pressure. 
The results are validated against more refined URANS simulations for selected cases.

\end{abstract}

\section{Introduction}
\label{sec:Intro}
The flow around bluff bodies is common in nature as well as in many engineering applications such as road vehicles and bridges. 
It is mainly characterized by a large vortex shedding and a low pressure region in the wake, which is the main source of the elevated drag. 
% Improvement of aerodynamic performance can contribute to reductions in power consumptions, an important objective for sustainable transport.
The wake dynamics can also cause serious structural vibrations and acoustic noise. 
These issues can be minimized or even mitigated with the use of passive or active flow control (AFC) devices. 
The current study aims to numerically optimize a passive and an active flow control device for an existing D-shaped model.

Various passive means for bluff body drag minimization have been investigated during the last decades. 
For example, Roshko \cite{Roshko1954} and Tanner \cite{Tanner1975} investigated the benefits of a splitter plates. 
In Park et al. \cite{Park2006}, a 33\% drag decrease over a generic bluff body was achieved using distributed tabs over the blunt trailing edge.
A different approach using the surface roughness to thickens the boundary layer and thereby to reduce the drag was presented by Whitmore and Naughton \cite{Whitmore2002}.
% For flow over a model with a square back, the most dominant is the fixed separation point at the blunt trailing edge, and the consequent large recirculation bubble in the wake.
Base cavities and boat tails have also been shown to significantly reduce the size of the separation bubble or shift it further away from the base (e.g. Balkanyi et al. \cite{Balkanyi2002} and Verzicco et al. \cite{Verzicco2002}).
Optimizations of such boat-tails \textit{without} cavity have been conducted, such as in Han et al. \cite{Han1992} for the rear-end shape of a vehicle-like body.
However, to the author's knowledge, no optimization has been performed on boat tail geometries \textit{with} cavity, where drag reductions are expected to be larger. 

Most existing approaches to minimize bluff body drag are passive,
which are restricted by design and practical considerations and cannot be `turned off' when not needed. 
Moreover, away from their design operating conditions, passive devices are known to have a lot of adverse effects.
 
Because of the afore-mentioned shortcomings, active flow control (AFC) systems became the more popular alternatives \cite{Cattafesta2011}. 
Since active systems require energy to operate, it is critical that the power gained through drag reductions and the
use of AFC to be as large as possible in comparison to the expanded power of the blown jet. 
This relation can be quantified by the power ratio  \cite{Pfeiffer2014},
\begin{equation}
	\frac{\Delta P}{P_0}  = \frac{\Delta c_D}{c_{D,ref}} - \frac{P_a}{P_0} \, ,
	\label{eq::PowerRatio}
\end{equation}
where $\Delta c_D = c_{D,ref} - c_D$ is the drag coefficient difference between the actuated and the unactuated state, 
$P_0$ is the power needed to overcome the drag without actuation, and $P_a$ the power consumed by the actuator.
The drag coefficient is defined as
\begin{equation}
 c_D=\frac{D}{\frac{1}{2} \rho_\infty u_\infty^2 H}
\end{equation}
where $D$ is the drag force, $H$ is the body height (see figure \ref{fig:Configurations} (a)), and the subscript $\infty$ denotes far field conditions.

Active actuators for drag minimization incorporate a wide range of devices.
% The power savings for Bearman's study is not reported, whereas these for Pastoor et al. reached 2.8.
One common approach uses the zero-net-mass flux (ZNMF) actuators.
The advantage of ZNMF actuation is that no additional supply of air is needed. 
Park et al. \cite{Park2013} used synthetic jet arrays at the roof edge of an Ahmed body. 
% They reported a drag reduction of $5.2 \%$. 
However, ZNMF actuators are not easily scalable and their authority diminishes at high Reynolds numbers.
A simple actuator is the rotating cylinder, as presented by Beaudoin et al. \cite{Beaudoin2006},
where a significant drag reduction through open and closed-loop control was reported.
Another mechanism is base blowing, which could be performed through distributed (e.g. Bearman \cite{Bearman1967}) or localized (e.g. Pastoor et al. \cite{Pastoor2008}) blowing.
A special variant to the base blowing is the Coanda actuator, which is simply accomplished by adding a curved surface at the blowing slit exit.
Several studies using the Coanda blowing on a bluff body are reported in the literature, such as Freund et al. \cite{Freund1994} and Englar \cite{Englar2001}.
Coanda blowing has two effects on the wake of bluff bodies. 
Firstly, the induced flow that is attached to the Coanda surface helps to directly increase base pressure through added mass flux towards the base. 
Secondly, the injected momentum through the highly turning flow entrains the approaching flow and thereby helps it negotiate its way around the corner, resulting in a much narrower wake. 
This decreased wake width is closely linked to further increase in base pressure.
Optimizations on the Coanda actuator are hard to find in the literature.
Manosalvas et al. \cite{Manosalvas2014} performed a parametric study on the momentum coefficient ($c_\mu$) and the Coanda radius ($r_C$) to determine the minimal power coefficient. 
The momentum coefficient characterizes the blowing intensity as 
\begin{equation}
	c_\mu=\frac{\mathbf{U}_j \dot{m}_j}{\frac{1}{2} \rho_\infty u_\infty^2 H}\,,
\end{equation}
where $\mathbf{U}_j$ is the average jet velocity across the slit exit and $\dot{m}_j$ is the estimated mass flow rate of the jet.
In this study, the parameter defining the Coanda actuator ($\mathbf{U}_j$, $r_C$, and the slit height $h$) are optimized such that the power ratio is minimized.

\section{Numerical Setup}
\label{sec:NumSetup}
The present investigations are based on two-dimensional numerical simulations of a generic D-shaped bluff body. 
The details of the model configuration (\S \ref{sec:Config}), of the numerical set-up (\S \ref{sec:RANS}) and of the optimization sequence (\S \ref{sec:OptSeq}) are presented in the following.

\subsection{Configurations}
\label{sec:Config}
\begin{figure}[]
\begin{center}
	\subfigure[]{\includegraphics[width=0.7\textwidth]{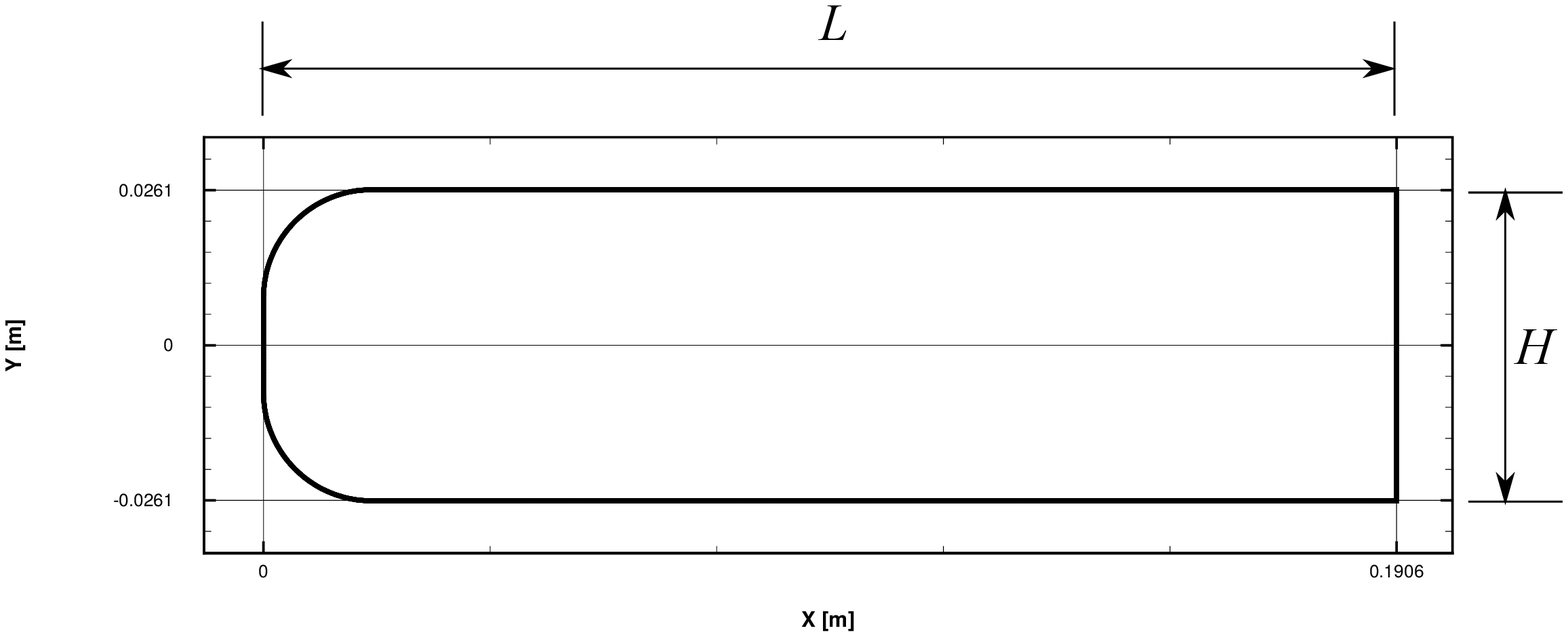}}
	\subfigure[]{\includegraphics[height=0.415\textwidth]{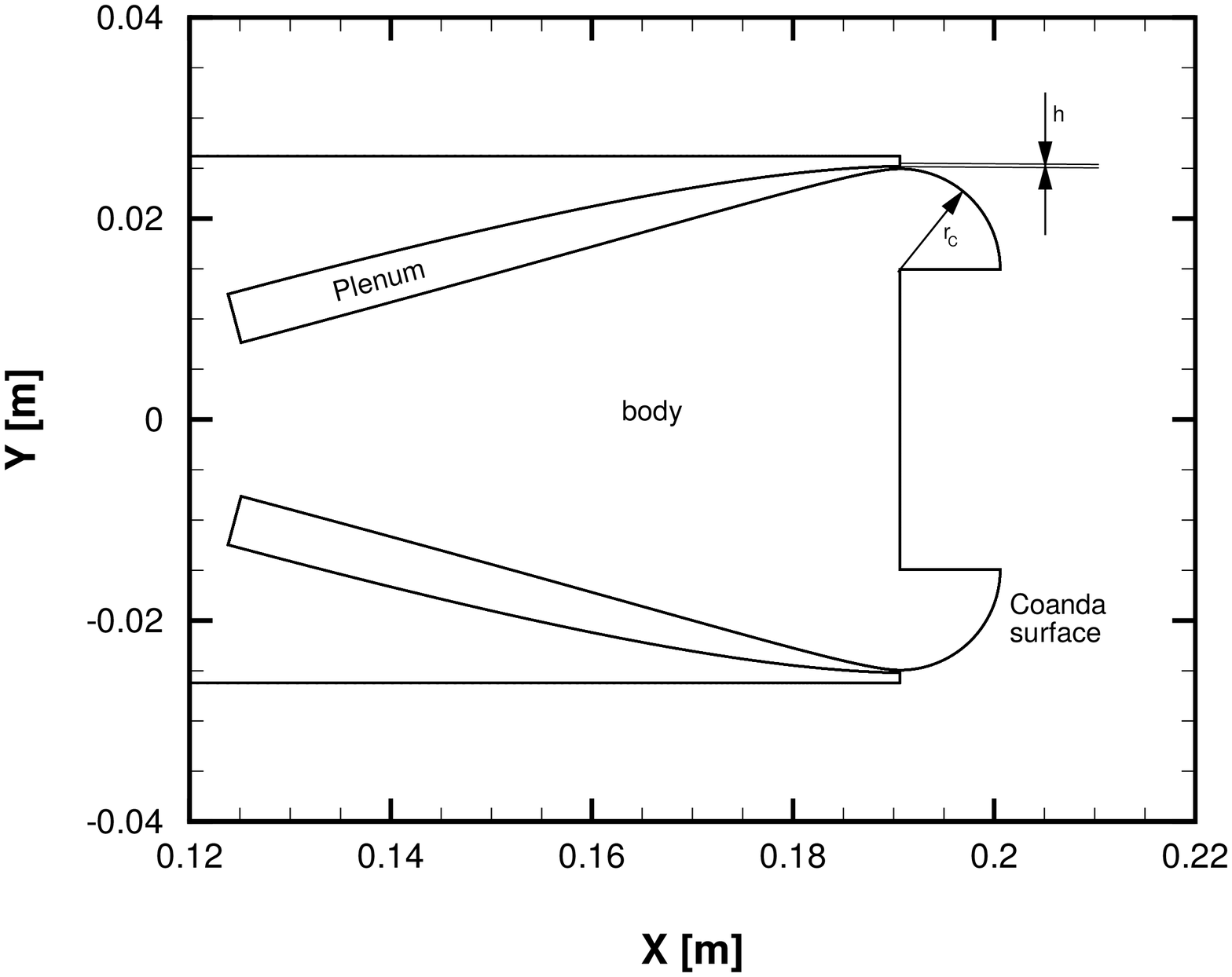}}
	\subfigure[]{\includegraphics[height=0.38\textwidth]{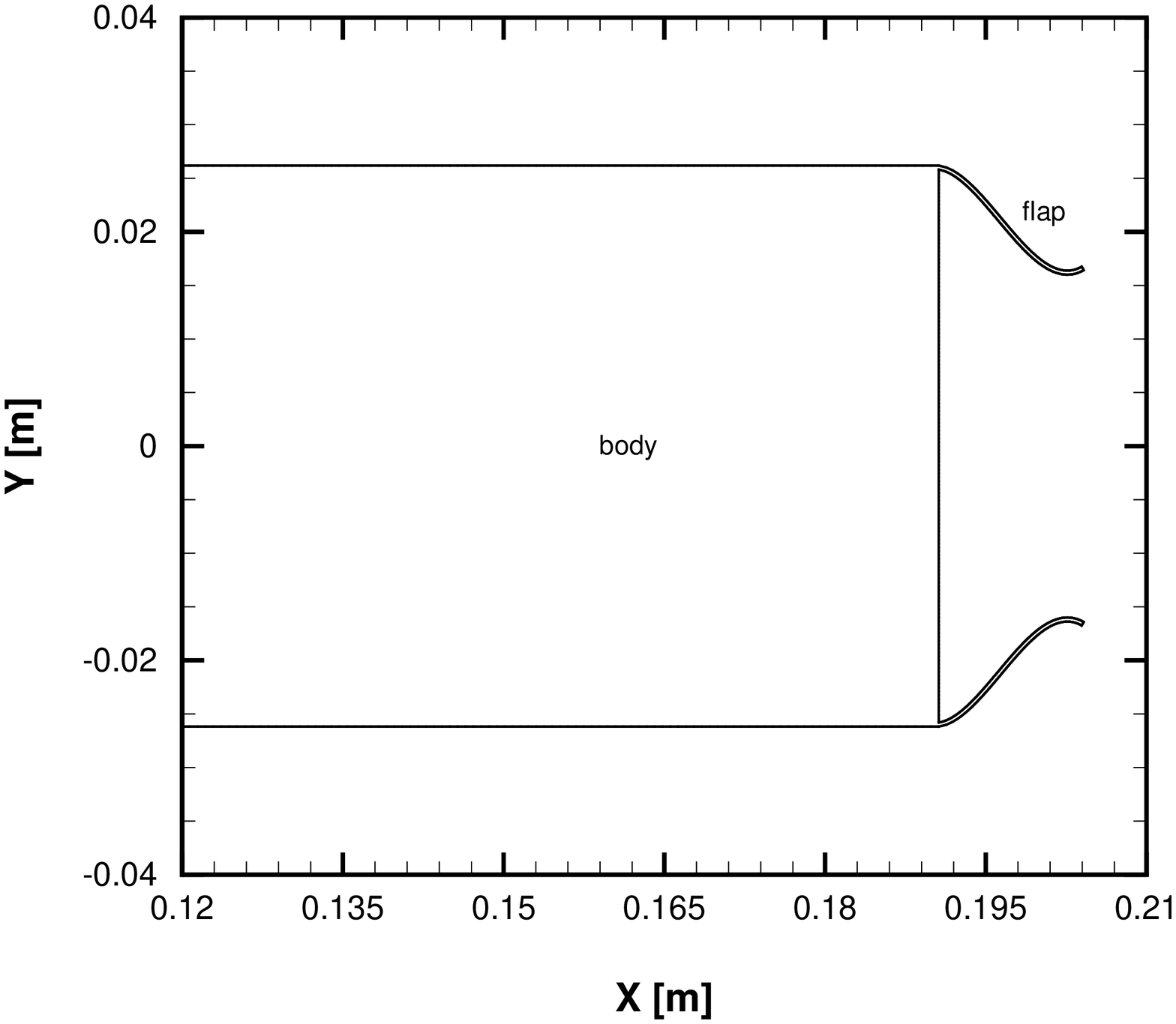}}
	\caption[D-shaped bluff body and details]{The different D-shaped bluff body configurations: (a) the reference configuration, (b) the configuration with the Coanda actuator, 
(c) and the configuration with the passive flaps.}
	\label{fig:Configurations}
\end{center}
\end{figure}

The base configuration, shown in figure \ref{fig:Configurations} (a), is a generic D-shaped bluff body similar to that of Pastoor et al. \cite{Pastoor2008}. 
The model geometry and dimensions match those of a similar experimental model currently being tested. 
Its length is $L = 0.1906$\,m whilst its height is $H = 0.0522$\,m. 
The leading edge of the body is rounded and the trailing edge is blunt. 
The two drag-minimizing devices, the Coanda actuator and the trailing edge flap, are added to the base of the reference configuration. 
One exemplary configuration of the active Coanda actuator and of the passive trailing edge flap are presented in figure \ref{fig:Configurations} (b) and (c), respectively. 

The Coanda actuation is geometrically defined through the slit height $h$ and the Coanda radius $r_C$, both illustrated in figure \ref{fig:Configurations} (b). 
The lip thickness above the blowing slit is fixed at $d_N = 1 \times 10^{-3}$\,m to match that of the experimental model.
The main condition to the Coanda surface is that the transition from the plenum to the quarter circle Coanda surface is tangent. 
Since the optimization process is based on a genetic algorithm, it is possible that the optimizer occasionally yields an individual with a set of unphysical parameters.
Therefore, the three defining variables (the slit height $h$, the Coanda radius $r_C$, and the plenum base velocity $u_{j,0}$) must be bounded.
A summary of the bounds is presented in table \ref{tab:OptCoandaSettings}.
The geometrical variables, the slit height $h$ and the Coanda radius $r_C$, are arbitrarily bounded (e.g. $r_C$ limited such that the model does not obtain a physical boat tail)
while keeping the solution space large enough.
The blowing velocity is also restricted from the plenum base at $u_{j,0}=3$\,m/s, such that the exit Mach number remains ${Ma} < 0.7$.
This is to limit the compressibility effects on the incompressible solver. 

The trailing edge flaps are parametrized as cubic polynomials
\begin{equation}
	f(x) = ax^3 + bx^2 + cx + d \, ,
\end{equation}
with $d = 0$ in order to keep the flap attached to the body.
The three remaining variables $(a, b, c)$ are bounded to avoid extreme (e.g. flap going very outwardly) or unphysical (e.g. the flaps intersecting each others) geometries.
Inspired by current European regulations \cite{EU2015} for allowable add-on devices, 
the length and thickness of the flaps is fixed at $l_{F} = 0.07 \times L = 1.334 \times 10^{-2}$m and $d_{F} = 0.002 \times L = 3.81 \times 10 ^{-4}$m, respectively.
Table \ref{tab:OptFlapSettings} summarizes all the imposed constraints.

\begin{table}[]
	\caption{Summary of the Coanda actuator geometric and blowing bounds}
	\begin{center}
	\begin{tabular}{lllll}
		\hline
		\textbf{Description} & & \textbf{Variable} & &\textbf{Bounds} \\ \hline
		Coanda radius & & $r_{C}$ & &$r_{C} \in [0.01, 10] \times 10^{-3}$\,m \\  
		Coanda jet slit height & & $h$ & & $h \in [0.01, 1.5] \times 10^{-3}$\,m \\ 
		Chamber inlet velocity & & $u_{j,0}$ & & $u_{j,0} \in [0.001, 3]$\,m/s \\ \hline
	\end{tabular}
	\end{center}
	\label{tab:OptCoandaSettings}
\end{table}
\begin{table}[]
	\caption{Summary of the flap geometric bounds}
	\begin{center}
	\begin{tabular}{lllll}
		\hline
		\textbf{Description} & & \textbf{Variable} & & \textbf{Interval} \\ \hline
		Flap length & & $l_{F}$ & & $l_{F} = 0.07 \times L = 1.334 \times 10^{-2}$\,m \\
		Flap thickness & & $d_{F}$ & & $l_{F} = 0.002 \times L = 3.81 \times 10 ^{-4}$\,m \\
		Cubic coefficient & & $a$ & & $a \in [-0.1, 0.1]$ \\ 
		Quadratic coefficient & & $b$ & & $b \in [-0.3, 0.3]$ \\ 
		Linear coefficient & & $c$ & & $c \in [-0.8, 0.8]$ \\ \hline
	\end{tabular}
	\end{center}
	\label{tab:OptFlapSettings}
\end{table}

All numerical simulations are performed at a Reynolds number $Re_H=U_{\infty} \cdot H/\nu_{infty}=5.35\cdot10^4$, where $\nu_{\infty}$ is the kinematic viscosity of the fluid. 
These flow parameters correspond to the wind tunnel conditions.

\subsection{RANS Simulations}
\label{sec:RANS}
The computational fluid dynamics (CFD) solver employed to perform the analysis is OpenFOAM (Open Field Operation and Manipulation).
The two-dimensional steady Reynolds-averaged Navier-Stokes (RANS) equations are solved using a finite volume approach. 
The discretization schemes are the second-order backward scheme for time, and a combination of first and second order for the convection, momentum, and Laplacian operators. 
Two turbulence turbulence models, Spalart--Allmaras and $k-\omega$ SST, are used for the Coanda actuator and for the flap geometry optimization, respectively.
The `downgrade' from the $k-\omega$ SST to the Spalart--Allmaras turbulence model for the Coanda configuration was due to computational cost considerations.

The mesh density is determined by means of a mesh convergence exercise.  
Three different grid densities for each configuration are tested. 
The medium grids, with $\approx,000$ and $\approx90,000$ points for the Coanda and the flap actuators, respectively, are chosen.
This represent a compromise between accuracy and computational cost.
It is worth to mention that the meshing process is repeated for every individual during the optimization process.
Therefore, the meshing macro in Pointwise${}^{\textregistered}$ had to be adaptable to the varying geometry.  

\begin{figure}[]
\begin{center}
	\subfigure[]{\includegraphics[height=0.405\textwidth]{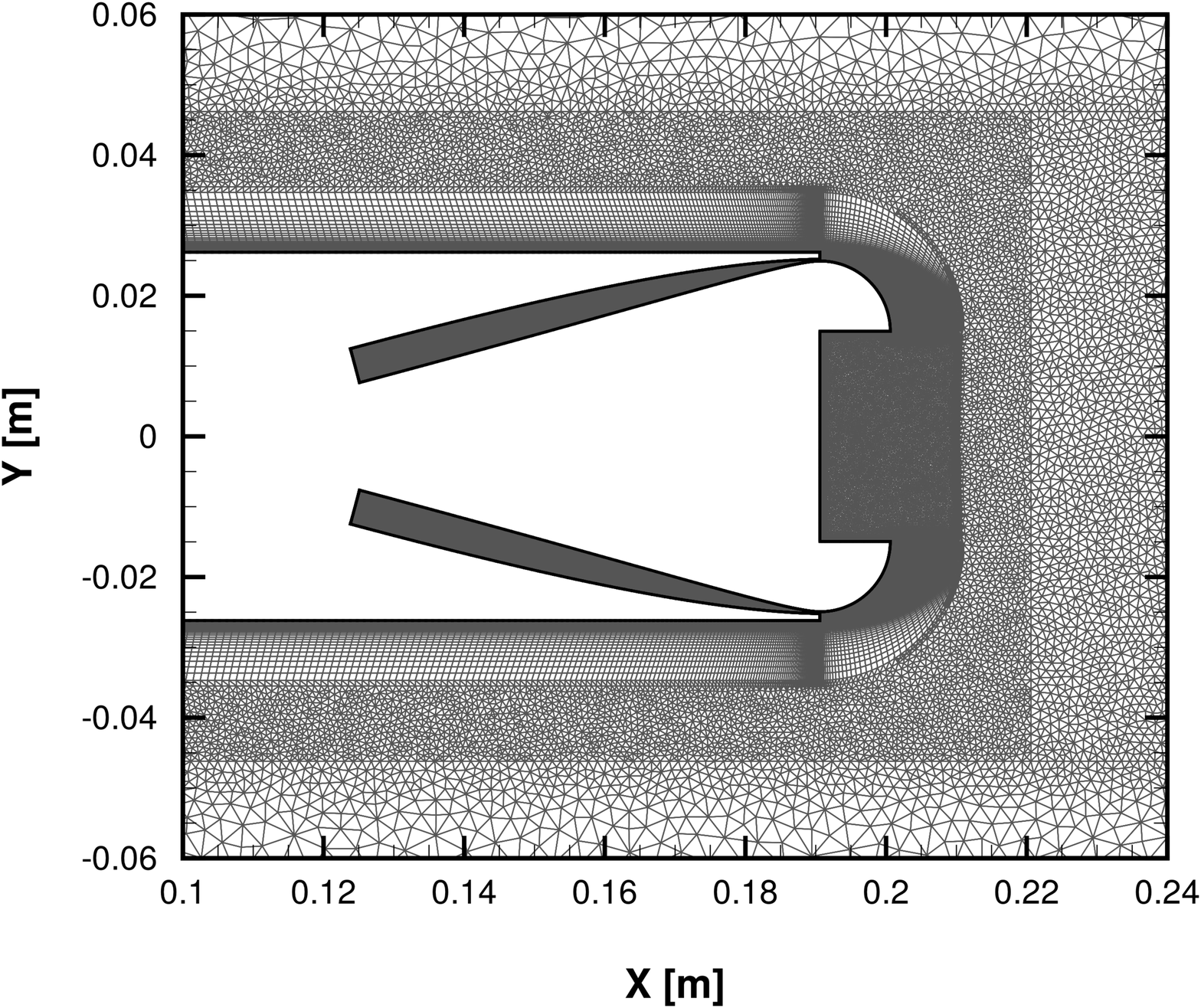}}
	\subfigure[]{\includegraphics[height=0.425\textwidth]{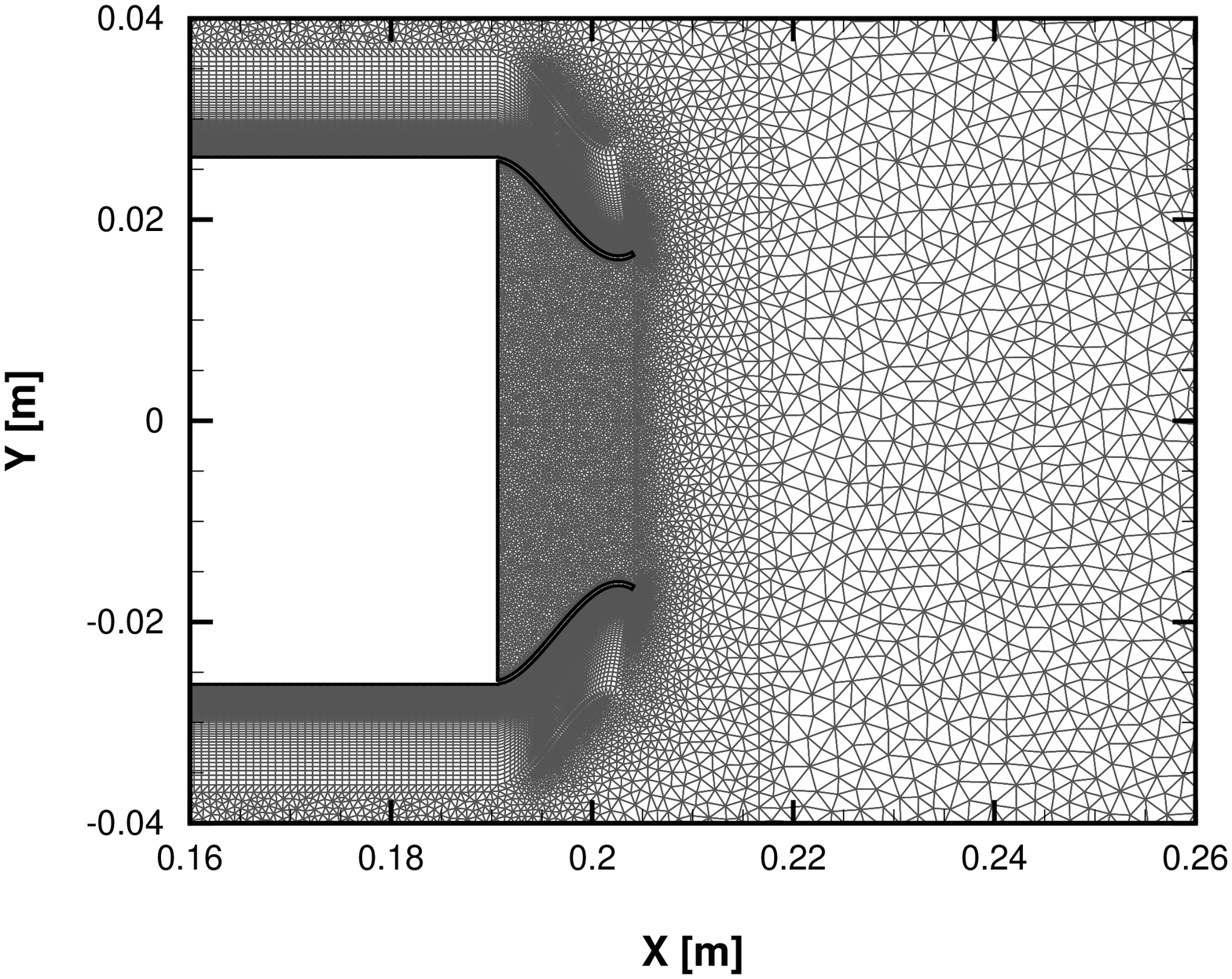}}
	\caption[A close-up of the meshes]{A close-up of the meshes for the (a) Coanda actuator, and the (b) trailing edge flap configurations.}
  \label{fig:Mesh}
\end{center}
\end{figure}

Both meshes are composed of a structured and an unstructured region, as the close-up in figure \ref{fig:Mesh} shows.
The inner structured mesh envelops the model to resolve the viscous boundary layer,
whereas the outer unstructured mesh has a C-block topology and extends 50 chord lengths in all directions.
The viscous sub-layer is resolved with $y_+<1$ everywhere over the airfoil surface. 
The slot lip is discretized by means of a local C-block topology.

In addition to the steady RANS simulations used for optimization, 
three URANS simulations using the highest mesh densities are performed.
The uRANS simulations are conducted for validation of three cases:
the refrence configuration, the optimal configuration with the Coanda actuators and the optimal configuration with the trailing edge flaps.
For brevity, the details of the validation URANS simulations are omitted.

\subsection{Genetic Algorithm}
\label{sec:GA}
There exist a wide range of optimization algorithms in the literature.
Due to their simplicity and robustness, most common optimizers are gradient-based. 
However, gradient-based methods are only efficient for small sets of independent variables \cite{Gregg1987}. 
For problems with a large number of variables that may be coupled, genetic algorithms (GA) are favored (e.g. Hajela \cite{Hajela1990}). 
Genetic algorithms have the added advantage of not getting stuck at local extrema. 
This feature helps them in finding global extrema more reliably, though convergence is not guaranteed. 
The details of genetic algorithm can be found in numerous publications (e.g. Duriez et al. \cite{Duriez2017}), 
but an overview is provided here.

Genetic or evolutionary algorithms form an important category of machine learning techniques inspired by the process of natural selection. 
A population of individuals, called a generation, compete to minimize a cost functional with the aim to propagate successful individuals to the next generation. 
An individual is a set of parameter values or a candidate solution to be optimized.
The initial generation is initially populated randomly.
Each individual is then assessed based on the evaluated cost functional. 
Individuals with a lower cost solution are more likely to proceed to the next generation. 
When evaluation of one generation is finished a new generation is created using various operations: elitism, replication, crossover and mutation.
In this study, the crossover, the replication, and the mutation rates are set to 80\%, 12\%, and 8\%, respectively.
Respectively, 20 and 25 generations for the Coanda actuator and the trailing edge flaps are conducted.

The genetic algorithm in this study is implemented using DAKOTA (Design Analysis Kit for Optimization and Terascale Applications) \cite{Dakota2015} toolbox from Sandia national laboratories. 
DAKOTA includes a vast library of optimization algorithms as well as uncertainty quantification and sensitivity analysis tools.
% A big advantage of DAKOTA is the ability to work with every possible application to perform the evaluations. 
In this work a single objective GA (\verb+soga+) of the JEGA library is used. 
% DAKOTA also allows to run evaluations in parallel which is crucial to the optimization task in this work, as it speeds up the whole optimization run dramatically.

\subsection{Optimization Sequence}
\label{sec:OptSeq}
% As shown in figure \ref{fig:OptSequence}, the optimization sequence consists of several inter-connected steps. 
The optimization sequence consists of several inter-connected steps. 
It is initialized by DAKOTA, which generates a set of variables characterizing the geometry as well as the actuation intensity. 
We recall that the variables are the polynomial coefficients $(a, b, c)$ for the flaps, and the Coanda geometric ($h$, and $r_C$) and blowing velocity at the plenum base $u_{j,0}$. 
These variables are then processed by Matlab to generate the corresponding shape, which is then transferred to Pointwise for meshing. 
The mesh is subsequently used by OpenFOAM for the RANS numerical simulations. 
After each simulation, the cost functional is calculated and then passed to DAKOTA which evaluates it and generates a new set of variables, and the cycle continues. 
For the passive device, the cost functional is the drag coefficient $c_D$, whereas for the active device it is the power ratio $\frac{\Delta P}{P_0}$.
% \begin{figure}[b]
% \sidecaption
% \includegraphics[width=0.6\textwidth]{figures/Evaluation.eps}
% \caption{Schematic illustration of the optimization sequence}
% \label{fig:OptSequence}
% \end{figure}

%
\section{Results}
\label{sec:Results}

\subsection{Coanda Actuator}
\label{sec:Coanda}
\begin{figure}[]
\begin{center}
	\subfigure[]{\includegraphics[height=0.38\textwidth]{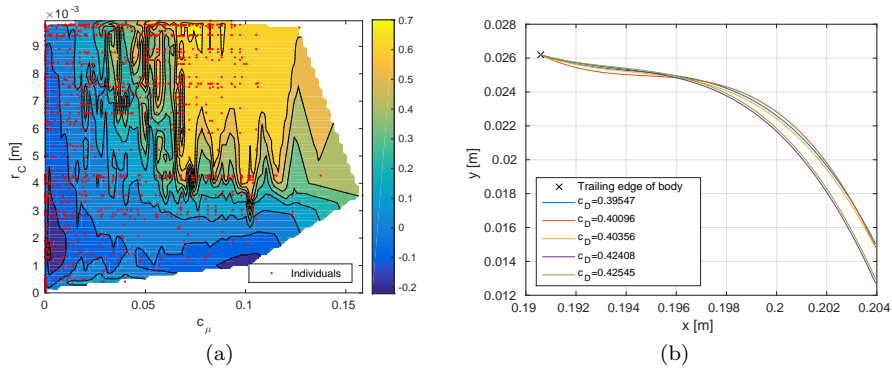}}
	\subfigure[]{\includegraphics[height=0.355\textwidth]{figures/ResFlapShortShapes.eps}}
	\caption[Final Results]{(a) The power ratio distribution over the $r_C-c_{\mu}$ range, and (b) the best 5 shapes for the trailing edge flaps.}
  \label{fig:Results}
\end{center}
\end{figure}

Figure \ref{fig:Results} (a) shows the power ratio distribution over the $r_C-c_{\mu}$ range for all tested individuals.
As the figure shows, there is a clustering of high power ratio solutions around $c_\mu \in [0.067, 0.077]$ and $r_C \in [9.2, 9.8] \times 10^{-3} \ m$. 
Within this region the power savings are $\approx 70 \%$, and the drag is drastically reduced by $\approx 82\%$ to $c_D =0.12$. 
The corresponding values of $h$ and $\mathbf{U}_j$ within this cluster are $h \in [3.7, 3.9] \times 10^{-4} \ m$ and $\mathbf{U}_j \in [33.7, 35.6] \ \frac{m}{s}$. 
Hence, the optimized and most robust combination can be chosen as: ${h} = 3.75  \times 10^{-4}$m, $r_C=9.5 \times 10^{-3}$m and ${\mathbf{U}_j} = 35.0$m/s.
% The results are comparable to those of Englar \cite{Englar2001}.

The first observation regarding the optimized parameters is the favorability of the optimizer towards larger Coanda radii.
This is understandable, as larger Coanda surfaces also contribute passively to drag reduction by acting as boat-tails and streamlining the wake.
A closer inspection of the flow around the optimal configuration reveals that the flow is only partially attached.
Here, the total size of the mean recirculation region is more streamlined than that of the base configuration: it is longer and thinner. 
This streamlining of the wake appears to have a similar effect to physical boat-tailing.
This suggests that the most energy-saving configuration does not require a complete suppression of the recirculation region.
Streamlining the wake appears to be the most energy efficient approach.

\subsection{Trailing Edge Flaps}
\label{sec:TE_Flaps}
Figure \ref{fig:Results} (b) shows the five best flap geometries determined by the GA.
The corresponding polynomial coefficients as well as the drag coefficients are listed in table \ref{tab:ResFlapShortShapes}.
As the table shows, the drag reduction is a respectable 41\% compared to the base configuration.
Surprisingly, none of the five geometries is tangent to the body's surface.
In fact the best flap geometries depict a small depreciation around $x\approx 0.193$ and an inflection point at $x\approx 0.195$.
Examining the flow around this depreciation reveals a small standing (or trapped) vortex,
which appears to aid guiding the flow inwardly.
This standing vortex is also observed in the more refined URANS simulations.

The drag reduction mechanisms are similar to those of the Coanda actuators.
The recirculation region behind the body is considerably elongated with higher pressure at the base.
The flaps' addition also creates smaller recirculation regions in the base cavity, 
which also act to increase the base pressure and to push the larger dead water region away from the body \cite{Barros2015}.
\begin{table}[]
	\caption{The polynomial coefficients of the five best trailing edge flaps}
	\label{tab:ResFlapShortShapes}
	\begin{center}
	\begin{tabular}{lllll}
	\hline
	\textbf{No.} & $\mathbf{a}$ & $\mathbf{b}$ & $\mathbf{c}$ & $\mathbf{c_D}$ \\ \hline
	1 & -0.0096	& 0.1044	& -0.5111 	& 0.3954 \\ 
	2 & -0.0122	& 0.1541	& -0.7249	& 0.4009 \\ 
	3 & -0.0096	& 0.1044	& -0.5302	& 0.4035 \\ 
	4 & -0.0096	& 0.0851	& -0.4283	& 0.4240 \\ 
	5 & -0.0096	& 0.0851	& -0.4102	& 0.4254 \\ \hline
	\end{tabular}
	\end{center}
\end{table}

\section{Conclusions}
Shape optimization of two drag reducing devices for a generic D-shaped bluff body is performed. 
The two devices are the Coanda actuator and the nonlinear trailing edge flap. 
Optimization is conducted using genetic algorithm DAKOTA coupled with OpenFOAM solver for the RANS simulations. 
The cost functional for the active actuator (Coanda) is the power ratio, whereas for the passive device (flap) it is the drag coefficient.
The optimization results are validated using URANS simulations (not shown).

Both optimized devices exhibit drastic decreases in drag. 
The best individuals of the trailing edge flaps yield drag reductions of $\approx 40 \ \%$ compared to the reference configuration. 
For the Coanda actuator, the drag reductions are as high as $\approx 82\%$ with a net power savings of $\approx 70\%$. 
In both cases the main effect leading to the drag decrease is the wake streamlining.
The URANS simulations revealed an additional effect which is the reduction or even suppression of vortex shedding in the wake.

%\begin{acknowledgements}
%\end{acknowledgements}


\bibliographystyle{spbasic}
\bibliography{STAB_Lit2.bib}

\begin{thebibliography}{22}
\providecommand{\natexlab}[1]{#1}
\providecommand{\url}[1]{{#1}}
\providecommand{\urlprefix}{URL }
\expandafter\ifx\csname urlstyle\endcsname\relax
  \providecommand{\doi}[1]{DOI~\discretionary{}{}{}#1}\else
  \providecommand{\doi}{DOI~\discretionary{}{}{}\begingroup
  \urlstyle{rm}\Url}\fi
\providecommand{\eprint}[2][]{\url{#2}}

\bibitem[{Adams et~al(2015)Adams, Ebeida, Eldred, Jakeman, Swiler, Stephens,
  Vigil, Wildey, Bohnhoff, Dalbey, Eddy, Hooper, Hu, Bauman, Hough, and
  Rushdi}]{Dakota2015}
Adams BM, Ebeida MS, Eldred MS, Jakeman JD, Swiler LP, Stephens JA, Vigil DM,
  Wildey TM, Bohnhoff WJ, Dalbey KR, Eddy JP, Hooper RW, Hu KT, Bauman LE,
  Hough PD, Rushdi A (2015) Dakota, a multilevel parallel object-oriented
  framework for design optimization, parameter estimation, uncertainty
  quantification, and sensitivity analysis: Version 6.3 user's manual. Sandia
  Technical Report SAND2014-4633, Sandia National Laboratories

\bibitem[{Balkanyi et~al(2002)Balkanyi, Bernal, and Khalighi}]{Balkanyi2002}
Balkanyi SR, Bernal LP, Khalighi B (2002) Analysis of the near wake of bluff
  bodies in ground proximity. In: ASME 2002 International Mechanical
  Engineering Congress and Exposition, pp 705--713

\bibitem[{Barros et~al(2016)Barros, Bor\'{e}e, Noack, Spohn, and
  Ruiz}]{Barros2015}
Barros D, Bor\'{e}e J, Noack BR, Spohn A, Ruiz T (2016) Bluff body drag
  manipulation using pulsed jets and {C}oanda effect. J Fluid Mech 805:422--459

\bibitem[{Bearman(1967)}]{Bearman1967}
Bearman PW (1967) The effect of base bleed on the flow behind a two-dimensional
  model with a blunt trailing edge. Aeronautical Quarterly 18:207--224

\bibitem[{Beaudoin et~al(2006)Beaudoin, Cadot, Aider, and
  Wesfreid}]{Beaudoin2006}
Beaudoin JF, Cadot O, Aider JL, Wesfreid JE (2006) Drag reduction of a bluff
  body using adaptive control methods. Physics of Fluids 18(8):085,107

\bibitem[{Cattafesta and Sheplak(2011)}]{Cattafesta2011}
Cattafesta LN, Sheplak M (2011) Actuators for active flow control. Annual
  Review of Fluid Mechanics 43(1):247--272

\bibitem[{{Directive (EU) 2015/719}(2015)}]{EU2015}
{Directive (EU) 2015/719} (2015) amending council directive 96/53/ec laying
  down for certain road vehicles circulating within the community the maximum
  authorised dimensions in national and international traffic and the maximum
  authorised weights in international traffic. Official Journal of the European
  Union 115:1

\bibitem[{Duriez et~al(2017)Duriez, Brunton, and Noack}]{Duriez2017}
Duriez T, Brunton S, Noack B (2017) Machine Learning Control -- Taming
  Nonlinear Dynamics and Turbulence. Fluid Mechanics and Its Applications,
  Springer

\bibitem[{Englar(2001)}]{Englar2001}
Englar RJ (2001) Advanced aerodynamic devices to improve the performance,
  economics, handling and safety of heavy vehicles. In: {SAE} Technical Paper
  Series 2001-01-2072

\bibitem[{Freund and Mungal(1994)}]{Freund1994}
Freund JB, Mungal MG (1994) Drag and wake modification of axisymmetric bluff
  bodies using coanda blowing. Journal of Aircraft 31(3):572--578

\bibitem[{Gregg and Misegade(1987)}]{Gregg1987}
Gregg R, Misegade K (1987) Transonic wing optimization using evolution theory.
  In: {AIAA} Paper 87-0520

\bibitem[{Hajela(1990)}]{Hajela1990}
Hajela P (1990) Genetic search - an approach to the nonconvex optimization
  problem. {AIAA} Journal 28(7):1205--1210

\bibitem[{Han et~al(1992)Han, Hammond, and Sagi}]{Han1992}
Han T, Hammond D, Sagi C (1992) Optimization of bluff body for minimum drag in
  ground proximity. {AIAA} journal 30(4):882--889

\bibitem[{Manosalvas et~al(2014)Manosalvas, Economon, Palacios, and
  Jameson}]{Manosalvas2014}
Manosalvas DE, Economon TD, Palacios F, Jameson A (2014) Finding
  computationally inexpensive methods to model the flow past heavy vehicles and
  the design of active flow control systems for drag reduction. In: 32nd {AIAA}
  Applied Aerodynamics Conference

\bibitem[{Park et~al(2006)Park, Lee, Jeon, Hahn, Kim, Kim, Choi, and
  Choi}]{Park2006}
Park H, Lee D, Jeon WP, Hahn S, Kim J, Kim J, Choi J, Choi H (2006) Drag
  reduction in flow over a two-dimensional bluff body with a blunt trailing
  edge using a new passive device. Journal of Fluid Mechanics 563:389--414

\bibitem[{Park et~al(2013)Park, Cho, Lee, Lee, and Kim}]{Park2013}
Park H, Cho JH, Lee J, Lee DH, Kim KH (2013) Aerodynamic drag reduction of
  ahmed model using synthetic jet array. {SAE} International Journal of
  Passenger Cars - Mechanical Systems 6(1):1--6

\bibitem[{Pastoor et~al(2008)Pastoor, Henning, Nock, King, and
  Tadmor}]{Pastoor2008}
Pastoor M, Henning L, Nock BR, King R, Tadmor G (2008) Feedback shear layer
  control for bluff body drag reduction. J Fluid Mech 608

\bibitem[{Pfeiffer and King(2014)}]{Pfeiffer2014}
Pfeiffer J, King R (2014) Linear parameter-varying active flow control for a
  {3D} bluff body exposed to cross-wind gusts. In: {AIAA} Paper 2014-2406

\bibitem[{Roshko(1954)}]{Roshko1954}
Roshko A (1954) On the development of turbulent wakes from vortex streets. NACA
  Report 1191

\bibitem[{Tanner(1975)}]{Tanner1975}
Tanner M (1975) Reduction of base drag. Progress in Aerospace Sciences
  16(4):369--384

\bibitem[{Verzicco et~al(2002)Verzicco, Fatica, Iaccarino, Moin, and
  Khalighi}]{Verzicco2002}
Verzicco R, Fatica M, Iaccarino G, Moin P, Khalighi B (2002) Large eddy
  simulation of a road vehicle with drag-reduction devices. {AIAA} journal
  40(12):2447--2455

\bibitem[{Whitmore and Naughton(2002)}]{Whitmore2002}
Whitmore SA, Naughton JW (2002) Drag reduction on blunt-based vehicles using
  forebody surface roughness. Journal of Spacecraft and Rockets 39(4):596--604

\end{thebibliography}

\end{document}